\documentclass[twocolumn,twoside,preprintnumbers,amsmath,amssymb,pacs]{revtex4}
\usepackage{epsfig}
\usepackage{graphicx}% Include figure files
\usepackage{dcolumn}% Align table columns on decimal point
\usepackage{bm}% bold math

\usepackage{fancyhdr}

\usepackage{pslatex}

\begin{document}

\title{{\Large Non-local effects at the onset of the chiral transition}}

\author{ Let\'\i cia F. {\sc Palhares}$^1$\footnote{leticia@if.ufrj.br}, 
Eduardo S. {\sc Fraga}$^1$\footnote{fraga@if.ufrj.br},
Takeshi {\sc Kodama}$^1$\footnote{tkodama@if.ufrj.br},
Gast\~ao {\sc Krein}$^2$\footnote{gkrein@ift.unesp.br} }
\affiliation{$^1$Instituto de F\'\i sica, 
Universidade Federal do Rio de Janeiro\\
C.P. 68528, Rio de Janeiro, RJ 21941-972, Brazil\\
$^2$Instituto de F\'\i sica Te\'orica, Universidade Estadual Paulista\\
Rua Pamplona 145, S\~ao Paulo, SP 01405-900, Brazil}

%\received{on 31 March, 2006}

%%%%%%%%%%%%%%%%%%%%%%%%%%%%%%%%%%%%%%%%%%%%%%%%%%%%%%%%%%%%%%%%%

\begin{abstract}
Inspired by analytic results obtained for a systematic 
expansion of the memory kernel in dissipative quantum mechanics, 
we propose a phenomenological procedure to incorporate 
non-markovian corrections to the Langevin dynamics of an 
order parameter in field theory systematically. In this note, 
we restrict our analysis to the onset of the evolution. As 
an example, we consider the process of phase conversion in 
the chiral transition.

%PACS numbers:

%Keyword:
\end{abstract}

%%%%%%%%%%%%%%%%%%%%%%%%%%%%%%%%%%%%%%%%%%%%%%%%%%%%%%%%%%%%%%%%%

\maketitle

\thispagestyle{fancy}
\setcounter{page}{0}

%%%%%%%%%%%%%%%%%%%%%%%%%%%%%%%%%%%%%%%%%%%%%%%%%%%%%%%%%%%%%%%%%

\section{Introduction}

For high enough values of temperature, strongly interacting 
matter should be in a quark phase due to asymptotic freedom.
For the last few years, relativistic high-energy heavy ion 
collisions have been providing valuable information on the 
new state of matter that seems to have been created according 
to recent data from experiments at BNL-RHIC \cite{QM2004}, 
even though its true nature is still uncertain. Moreover, 
comparison between temperature estimates from data and results 
from Lattice QCD apparently shows that the critical temperature 
($T_c \sim 150$ MeV) for the chiral transition has been exceeded 
\cite{Laermann2003}.

As the deconfined plasma created in a heavy-ion collision expands, 
it cools down, undergoing a phase transition (or a rapid 
crossover) back to hadronic matter, as well as breaking
the approximate chiral symmetry.  Results from CERN-SPS 
and BNL-RHIC feature an explosive behavior in the hadronization 
process and seem to favor a very fast spinodal decomposition as 
the mechanism of phase conversion. Effective chiral 
field theory models for QCD also point to an explosive spinodal 
decomposition scenario \cite{explosive,Scavenius:2001bb}. 

Under these conditions, since relevant time scales are not 
so clearly separated (as, for instance, in the case of the 
primordial quark-hadron transition), the competition between several 
dynamical phenomena will significantly affect the phase conversion
process. In this context, fluctuations and non-local effects will 
play a major role in the dynamics of the order parameter. Therefore, 
a linear response non-equilibrium field theory approach will provide 
a highly non-local Langevin equation for the evolution of the chiral 
condensate \cite{EffQFT}, and the memory kernel details will presumably 
be important in the determination of the physical time scales for 
the whole conversion process \cite{ESF&GK2005}. Most likely one 
will find interesting transient effects and, eventually, a hierarchy 
of time scales.

Since the structure of memory integrals and colored noise that 
appear in field theory, even for relatively simple effective 
models, is often rather complicated, one has to resort to 
numerical simulations to obtain exact results 
\cite{numerics-qft}. Analytic results, 
which could bring some qualitative understanding of the 
mechanisms involved, can only be achieved within systematic 
approximations \cite{Qnonmarkov}. 

In this paper, we propose a phenomenological procedure to incorporate 
non-markovian corrections to the Langevin dynamics of an order parameter 
in field theory systematically. We follow a path inspired by analytic 
results obtained for a systematic expansion of the memory kernel in 
dissipative metastable quantum mechanics, a much simpler problem which 
we use as a toy model. To avoid some technical complications, we restrict 
our analysis to the onset of the evolution, {\it i.e.}, to very early times. 
Moreover, for simplicity, we intentionally neglect inhomogeneity effects 
and the noise contribution, although we know they will be important 
and should be incorporated \cite{bruno,Fraga:2005wd}. These issues, 
as well as a detailed comparison to exact (numerical) results, will 
be addressed in a future publication \cite{Qnonmarkov}. As an application, 
we consider the process of phase conversion in the chiral transition, 
comparing our results to those from a markovian Langevin evolution 
\cite{ESF&GK2005}. In view of the caveats mentioned above, our results 
for the initial dynamics of the order parameter should be taken 
as a hint to the complex transient regime that can only be unveiled 
by a more rigorous approach.

The paper is organized as follows. Section II presents the method 
in the framework of non-markovian dissipative quantum mechanics, 
giving special attention to the features that will be important 
in the phenomenological extension to field theory. The early-time 
dynamics of the chiral condensate with the first phenomenological 
non-local corrections is discussed in Section III. Section V contains 
our conclusions and outlook. 

%%%%%%%%%%%%%%%%%%%%%%%%%%%%%%%%%%%%%%%%%%%%%%%%%%%%%%%%%%%%%%%%%
%
\section{Non-markovian dissipative metastable quantum mechanics}

The development of a systematic procedure to take non-local 
effects into account should have as starting point a careful 
determination of the domain of validity of the local
approximation. However, as pointed out above, the structure of 
memory kernels that come about in non-equilibrium field-theoretic 
approaches is rather intricate. This may turn the determination
of the domain of validity
into an excessively complex task. To isolate
the analysis of the role of non-locality, we consider a much 
simpler case in quantum mechanics (QM), where the problem can be solved
in a detailed and controlled fashion. 

Our QM toy model is constituted by a particle, with coordinate $q$ and
momentum $p$, interacting linearly with a heat bath which introduces 
dissipation in the dynamical evolution. The reservoir is modeled in 
the usual Caldeira-Leggett fashion \cite{Caldeira&Leggett}. The 
hamiltonian is thus given by:
\begin{eqnarray}
H &=& \frac{p^2}{2 M}+V(q) + H_R + \theta(t)q  \sum_k c_k R_k
\, ; \label{H}  \\
%
%H_0 + H_R + H_{I}(t) \, ; \label{H}  \\
%H_0 &=& \frac{p^2}{2 M}+V(q) \, , \nonumber \\ 
H_R &=&  \sum_k \left(\frac{p_k^2}{2m}+\frac{1}{2}m \omega_k^2 R_k^2
\right)
\, ,
%H_{I}(t)&=&\theta(t)q  \sum_k c_k R_k \, , \nonumber
\end{eqnarray}
where the sum over the heat bath modes is limited by a ``cutoff'' $\Omega$,
that corresponds to the maximum frequency for the interaction of the 
system with the medium. We also assume the system to be suddenly 
immersed into the reservoir at $t=0$ (which represents a rough 
simulation of the system ``quench'' in the chiral transition), 
so that the approximation of an initially uncorrelated density 
matrix is reasonable:
\begin{equation}
\rho (t \le 0)=\left( |0\rangle\langle 0| \right)_q \otimes 
e^{-\beta H_R} \, ,
\label{initial}
\end{equation}
where $T=1/\beta$ is the reservoir temperature.

The evolution towards equilibrium is studied within the 
Schwinger-Keldysh closed-time path framework \cite{SK}.
Integrating over the bath variables, we obtain the effective
action. After performing a Wigner transform to separate the
classical coordinate, $Q(t)$, from the fluctuations (associated 
with a Gaussian noise $\xi(t)$) the equation of motion assumes
the following semiclassical generalized Langevin form:
\begin{eqnarray}
M \ddot{Q}+ \overline{V}'\left( Q \right) + 
\frac{2 \eta}{\pi} \int^{t}_{0} dt' ~
\frac{\sin \left[ \Omega \left( t-t' \right) \right]}{t-t'} ~\dot{Q} 
\left( t' \right) &=& \xi \left( t \right) \nonumber \\
\langle \xi \left( t \right) \xi \left( t' \right) 
\rangle  =  \frac{2\eta T}{\pi} 
\frac{\sin \left[ \Omega \left( t-t' \right) \right]}{t-t'} \, &,& \label{eq1}
\end{eqnarray}
where $\eta$ is a phenomenological dissipation parameter encoding information
about the heat bath and
 $\overline{V}'\left( Q \right)$ is the derivative with respect
to $Q$ of the modified particle potential, 
\begin{equation}
\overline{V}'\left( Q \right) \equiv V'\left( Q \right) - 
\left[2 \eta\Omega/\pi\right]~Q  \, .
\end{equation}
In the markovian limit, this correction represents an ultra-violet
divergence that is associated with the unphysical situation in which
the particle interacts with all frequencies of the heat bath. 
Usually, in the context of quantum dissipative systems, this
alteration in the potential is ignored, under the argument
that the interest resides on the modifications brought about by dissipation
and not by medium renormalization effects. In quantum field theory, however,
the ``particle'' and the reservoir are deeply entangled, corresponding
to different Fourier modes of the same field (refer to \cite{Qnonmarkov} 
for further discussion). Since we are ultimately interested in the 
chiral transition case, we compare in this analysis the evolutions 
with and without the modification in the potential. 

The limit in which $\Omega$ is strictly infinite yields a local 
dissipation term in Eq. (\ref{eq1}). Therefore, the non-markovian 
contributions arise from the assumption that $\Omega$ is still 
large, but finite. As one goes down in $\Omega$ one should, 
in principle, include more and more non-local corrections. 
Taking advantadge of the localized profile of the kernel that appears
in Eq. (\ref{eq1}) and assuming that the system will eventually thermalize 
in a stable vacuum, we can derive the following series for the memory integral 
(for details, see \cite{Qnonmarkov}):
\begin{equation}
M \ddot{Q}+ \overline{V}'\left( Q \right) + 
\frac{2 \eta}{\pi} \sum^{\infty}_{n = 0} 
\frac{J_{n}(\Omega t)}{n!} 
~\frac{Q^{(n+1)} \left( t \right)}{\Omega^n} 
= \xi \left( t \right) \,, 
\label{eq2}
\end{equation}
where the time-dependent coefficients $J_{n}(\Omega t)$ contain 
information about the specific form of the original kernel. It should 
be noticed that the first term corresponds to the usual 
markovian approximation, in the limit $\Omega \to \infty$. 

The oscillatory
behavior of the integrand in Eq. (\ref{eq1}) is inherited by the $J_n$'s.
Nevertheless, $J_0$ and $J_1$ have definite signs: $J_0>0$ and $J_1\le 0$.
However, for $ n \ge 3 $, one can show that the signs of those
coefficients change according to the time at which it is being evaluated. 
Even though the amplitudes of the associated terms decrease, rendering these 
oscillations irrelevant for long times, the transient dynamics will
be greatly affected. 

We remark that the localized profile of the memory kernel is the 
essential feature for the development of
the non-local series described above. The oscillatory characteristic 
of the $J_n$'s is a specificity of the function $\sin (x)/x$ and its 
consequences do not encode modifications due to non-locality. Therefore, 
the alternating signs should not be interpreted as intrinsic 
memory effects. As a matter of fact, the essential feature being 
added here, as compared to the markovian case, is a {\it non-zero} 
correlation for times separated by an interval $\Delta t \lesssim 1/\Omega$. 
For $\Omega$ large enough, higher-order contributions in Eq. (\ref{eq2})
become more and more negligible and the series can be truncated. This
procedure systematically incorporates the effects from correlations 
between times within this interval. 

Now we analyze the non-local Langevin evolution dictated by 
Eq. (\ref{eq2}) for the double-well potential, namely for 
$V'(Q)=4\lambda Q^3-M\omega_0^2 Q$, at early times. In this 
regime, the equation is linear. For simplicity, we also 
approximate the coefficients by constants.
Keeping terms up to the third-order time derivative, one has:
\begin{eqnarray}
\overline{M} \ddot{Q}- \left[M\omega_0^2 + 2 \eta\Omega/\pi\right]~Q
 +\eta ~\dot{Q}+
\eta_{3} \stackrel{...}{Q} = \xi \left( t \right) \label{eqQM}\\
\overline{M}=M\left( 1+\frac{2}{\pi}\frac{\eta}{M\Omega} J_1 \right)
\quad\, ,\quad
\eta_3=\frac{2}{\pi}\frac{\eta}{\Omega^2} J_2 
\end{eqnarray}

\begin{figure}
\begin{center}
\includegraphics[width=8cm]{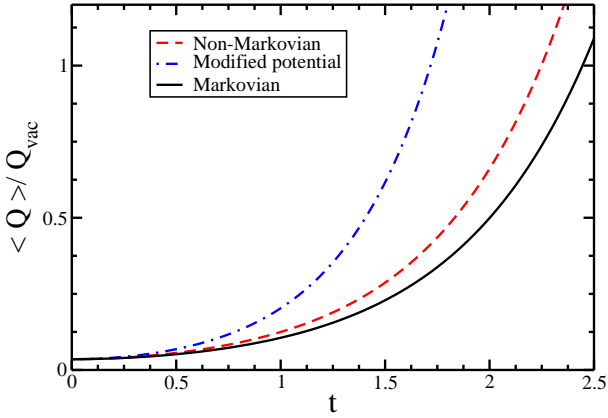}
\caption{The average of $Q$ in units of its vacuum value  
as a function of time for $M=\omega_0=\eta=2$ in arbitrary units
and $\langle Q(0)\rangle =0.04~Q_{vac}$. 
The dashed curve corresponds to the non-markovian case with 
$\Omega=2~\Lambda\approx3.852$, while the dot-dashed one is the 
non-local solution including the potential renormalization for the 
same value of $\Omega$. }
\label{figQM}
\end{center}
\end{figure}

The solution of Eq. (\ref{eqQM}) is obtained through a straightforward 
calculation, using the Laplace transform. Results for the average 
$\langle Q\rangle$ as a function of time are displayed in Fig. \ref{figQM},
for $ \langle \dot Q(0)\rangle= \langle \ddot Q(0)\rangle = 0 $.
One can see that both memory and potential corrections tend to accelerate 
the initial decay process towards the minimum at $Q=Q_{vac}$. 
It is clear that the modification of the potential introduces the 
major effect, remarkably reducing the time scale involved in the decay 
process. Although this correction may be ignored in the analysis of 
dissipative phenomena in QM, as discussed previously, it must be taken 
into account in the field theory case. In the latter, the medium 
renormalization of the effective potential should play a significant 
role in the phase transition process.

The fact that not only the medium modification of the potential 
but also the memory correction tend to speed up the decay of 
the particle seems surprising at first sight. In fact, one 
expects the latter to slow down the rollover process. The behavior 
observed in Fig. \ref{figQM} is certainly a transient effect and 
should be washed out in the long run by dissipation and non-linear 
effects \cite{Qnonmarkov}. Nevertheless, it points to the possibility 
of a non-trivial transient regime at early times. We will return to 
this point in the next section.

%%%%%%%%%%%%%%%%%%%%%%%%%%%%%%%%%%%%%%%%%%%%%%%%%%%%%%%%%%%%%%%%%
%
\section{Chiral transition early-time dynamics}

Inspired by the results obtained above for the QM toy model, 
we can construct a phenomenological generalization to the 
case of field theory. In particular, we consider the case 
of the chiral transition. The effective potential, 
$U_{eff}(\phi,T)$, for the chiral condensate $\phi(\vec{x},t)$ 
can be obtained, for instance, from an effective chiral model 
as described in Ref. \cite{Scavenius:2001bb}. 

The local form of the Langevin dynamics for the chiral condensate, 
which plays the role of a non-conserved order parameter in a 
Landau-Ginzburg approach, can be derived under convenient approximations 
within the framework of non-equilibrium field theory \cite{EffQFT}, 
and is given by
\begin{equation}
\Box~\phi+\Gamma(T) ~\frac{\partial \phi}{\partial t}+U'_{eff}(\phi,T)
=\xi(\vec{x},t) \, .
\label{eq3}
\end{equation}
The second term in Eq. (\ref{eq3}) corresponds to markovian dissipation
and the right-hand side represents a gaussian, white stochastic force, 
both attained, in principle, after drastic approximations of complicated 
memory kernels \cite{EffQFT}. Some of us have recently analyzed the 
role of dissipation in the hadronization of the quark-gluon plasma 
under these conditions \cite{ESF&GK2005}.

The inclusion of non-local effects is done, phenomelogically, by inserting
higher-order time derivatives of $\phi(\vec{x},t)$ into Eq. (\ref{eq3}):
\begin{equation}
\Box~\phi+\Gamma ~\frac{\partial \phi}{\partial t}+
\Gamma_2 ~\frac{\partial^2 \phi}{\partial t^2}+
\Gamma_3 ~\frac{\partial^3 \phi}{\partial t^3}+U'_{eff}(\phi,T)
=\xi(\vec{x},t) \, . \label{eq4}
\end{equation}
Trying to follow the scale hierarchy that arises naturally in the series 
expansion in QM, we choose the form of the coefficients accompanying 
these new terms. In addition to being proportional to the phenomenological 
temperature-dependent dissipation coefficient $\Gamma$, which can be 
estimated using kinetic theory, the $\Gamma_n$'s
are associated with increasing inverse powers of a large scale $\Omega$: 
$\Gamma_n(T)=(-1)^n\Gamma/\Omega^{n-1}$. Here, 
$\Omega$ is the parameter that controls the revelance of non-locality
and imposes the intended hierarchy between the contributions of different 
orders of time derivatives. For the above truncation of the series that 
presumably appears in Eq. (\ref{eq4}) to be valid, $\Omega$ must be large 
(although finite) compared to the temperature $T$, the natural scale of 
the system.

To obtain analytical solutions of Eq. (\ref{eq4}), one can resort to 
a sixth-order polynomial fit of $U_{eff}(\phi,T)$ \cite{ESF&GK2005} 
and investigate the short-time dynamics. Since we consider the 
explosive spinodal decomposition scenario, the potential is assumed 
to be time-independent. Its form is fixed at the spinodal temperature 
($T_{sp}\sim 108$ MeV), and the system is initially slightly displaced 
from the inflexion point ($\phi_0 \sim 0.162~ T_{sp}$).

For short times, $U'_{eff}$ can be approximated by its linear form around
$\phi_0$. We treat, for simplicity, the dynamics of the
mean value of the field over the stochastic noise for a homogeneous
initial condition.
Therefore, Eq. (\ref{eq4}) becomes equivalent to an infinite set
of linear ordinary differential equations: each Fourier mode of the field 
evolves according to an analogue of Eq. (\ref{eqQM}). When one examines 
the solution averaged over space, one verifies that the only relevant 
mode is $k=0$. Under these conditions, the field theory 
case is reduced to a QM problem for the Fourier component 
$\phi_{k=0}\equiv\varphi$:
\begin{eqnarray}
\ddot\varphi+a~\dot\varphi+b~\stackrel{...}\varphi+c~\varphi=0
\, ,\quad \quad \quad \quad \quad \label{eqest}\\
a=\frac{\Gamma}{1-\Gamma/\Omega}; ~
b=\frac{\Gamma}{\Omega^2\left(1-\Gamma/\Omega\right)}; ~
c=\frac{-U''_{eff}(\phi_0)}{1-\Gamma/\Omega} \, . \label{coeffs}
\end{eqnarray}

\begin{figure}
\begin{center}
\includegraphics[width=8cm]{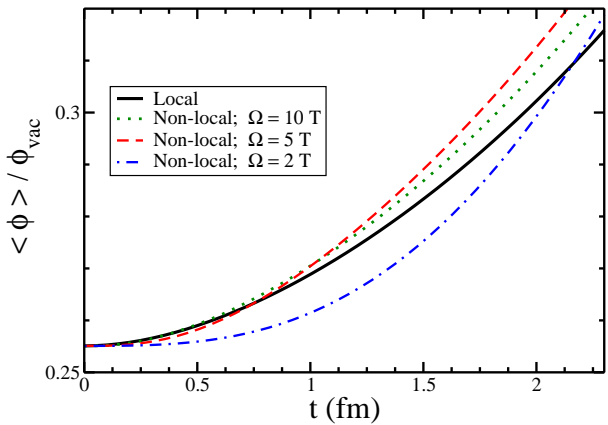}
\caption{The evolution at $T=T_{sp}$ of the average value of the order 
parameter over noise and space in units of its vacuum value. In addition 
to the local evolution (full line), three curves with non-markovian 
corrections are shown: $\Omega/T = 10$, $5$ and $2$.}
\label{figTQC}
\end{center}
\end{figure}

In Fig. \ref{figTQC}, we present results for the initial dynamics
of the order parameter $\phi(\vec{x},t)$ subject to
a non-markovian Langevin evolution. The behavior at times 
larger than those shown in the figure reproduces qualitatively 
the QM case, Fig. \ref{figQM}.

The existence of a competition between effects from different 
memory corrections is illustrated
in Fig. \ref{figTQC}. Although these results are valid only for
short times, the study of these curves enlightens the role 
of each term in the dynamics of the phase conversion.

The early influence of non-local dissipation is to retard the phase 
transition process. At extremely short times,
the lower-order time derivatives are subdominant, so that 
$\varphi\approx e^{\lambda t}$ with $\lambda\approx (b/a)^{1/3}
\propto \Omega^{2/3}$. Thus, the outcome of smaller values of $\Omega$
will be a greater delay in the initial evolution. On the other hand, for 
intermediate times, the modification brought about by the third-order 
derivative is overcome by the non-local correction of the potential. 
In this case, as $\Omega$ increases, the exponent $\lambda$ will decrease,
$\lambda\approx b^{1/2}$, and the curves tend to the markovian one.

Features that were neglected in this study should play a major
role in the transition dynamics, possibly washing out those transient
effects observed here. We will discuss the main shortcomings of the 
method in our conclusion.

%

%%%%%%%%%%%%%%%%%%%%%%%%%%%%%%%%%%%%%%%%%%%%%%%%%%%%%%%%%%%%%%%%%%%%%%%%%%
% 
\section{Conclusions and Outlook}

We constructed a phenomenological generalization of the Langevin 
dynamics experienced by an order parameter in field theory, accounting
for non-markovian correlations in a systematic fashion. We used as inspiration
a toy model in dissipative metastable QM, where the domain of validity of the local
approximation was identified, allowing us to derive a 
derivative expansion of the memory kernel. In this series, the first term 
corresponds to the usual markovian assumption and higher-order time 
derivatives are accompanied by increasing powers of the scale $1/\Omega$, 
associated with the maximum time interval in which correlations exist.

It should be noticed that this systematic expansion method developed can not 
be applied to any memory kernel. In fact, the kernel must have a localized
profile, so that correlations between sufficiently distant times vanish. 
This is in general not true when one considers a semiclassical expansion 
around a non-trivial background as, for instance, in the case of dynamical 
viscosity in nucleation \cite{dynnucleation}.

Results presented for the onset of the non-markovian evolution of the chiral 
condensate neglecting stochastic noise effects revealed an intricate transient
structure caused by the competition between different contributions. 
The influence of non-locality at extremely short times
is to delay the transtion process through a dissipation term
proportional to a higher-order derivative of the field.
At intermediate times, non-markovian modifications of
the potential dominate the evolution, accelerating it.

In this work, we identified the role of each non-local correction in 
a very simple model for the early dynamics. The complete evolution will be
studied numerically in a future publication \cite{Qnonmarkov}. 
In a more realistic approach, however, several other 
phenomena should be taken into account and an even more complex balancing 
will dictate the evolution towards the new vacuum state. The inclusion of 
the complete non-linear potential, for instance, would accelerate the 
early evolution and slow down the process for larger times, smoothening or 
even erasing the initial modifications observed above. The effect of stochastic
noise should be considered, as well as inhomogeneities in the effective
potential. Not only temporal, but also spatial non-locality should be
taken into account. All these aspects could, in principle, affect 
significantly the time scales of the phase conversion.

To obtain reliable quantitative results in the field theory case, one
must firstly derive this phenomenological non-local Langevin equation
through non-equilibrium quantum field theory. In this procedure, the
medium renormalization of the potential (which provides significant corrections
in the QM case) would be incorporated naturally and the scale $\Omega$ could
be identified and, possibly, estimated.  

%%%%%%%%%%%%%%%%%
\acknowledgments
This work was partially supported by CAPES, CNPq, FAPERJ, FAPESP and FUJB/UFRJ.

%%%%%%%%%%%%%%%%%%%%%%%%%%%

% FIGURES

%\begin{figure}[htbp]
%\begin{center}
%\includegraphics[width=8cm]{}
%\caption{}
%\end{center}
%\end{figure}

%FIGURES IN BOTH COLLUMNS
%\begin{figure*}[htbp]
%\begin{center}
%\includegraphics[width=8cm]{}
%\caption{}
%\end{center}
%\end{figure*}

%EQUATION IN BOTH COLUMNS
%\begin{widetext}
%\begin{equation}
%\end{equation}
%\end{widetext}

\end{document}